\documentclass[10pt,reqno,final]{article}

\usepackage{amsmath,amsfonts,amssymb,amsthm,version}
\usepackage{mathrsfs,fancybox,pifont}
\usepackage{graphicx}
\usepackage[colorlinks=true,linkcolor=blue,citecolor=blue,urlcolor=blue]{hyperref}
\usepackage{natbib}
\usepackage[notcite,notref]{showkeys}
\usepackage{color}
\usepackage{subfigure,multirow}
\usepackage{epstopdf}
\usepackage{cases}
\usepackage{mathtools}
\usepackage{algorithm,algorithmic}
\usepackage{authblk}

\usepackage{listings}
\usepackage{booktabs}
\usepackage{titlesec}
\titleformat{\chapter}[display]{\normalfont\huge\bfseries}{\chaptertitlename\ \thechapter}{0.8\baselineskip}{\Huge}
\titlespacing*{\chapter}{0pt}{0.8\baselineskip}{0.5\baselineskip}
\titleformat{\section}{\normalfont\Large\bfseries}{\thesection}{0.5\baselineskip}{}
\titlespacing*{\section}{0pt}{0.5\baselineskip}{0.5\baselineskip}
\usepackage{wrapfig}
\usepackage{multirow}
\usepackage{caption}

\allowdisplaybreaks

\usepackage[a4paper,left=3cm,right=3cm,top=2.5cm,bottom=2.5cm]{geometry}

\numberwithin{equation}{section}
\numberwithin{figure}{section}
\numberwithin{table}{section}

\theoremstyle{plain}

\usepackage{chngcntr}

\title{\textbf{Enhancement of the Prefiltered Rotationally Invariant Non-local PCA Algorithm for MRI}}

\author[1]{Shiao Li\thanks{2311110784@stu.pku.edu.cn}}

\affil[1]{Institute of Medical Technology, Peking University Health Science Center, Beijing 100191, China}

\date{}

\begin{document}
\maketitle
\counterwithout{figure}{section}
\counterwithout{table}{section}

\begin{abstract}

    Magnetic resonance imaging (MRI) is a non-invasive medical imaging technique offering high-resolution 3D images and valuable insights into human tissue conditions.
    Even at present, the refinement of denoising methods for MRI remains a crucial concern for improving the quality of the images. This study aims to enhance the prefiltered rotationally invariant non-local principal component analysis (PRI-NL-PCA) algorithm. 
    This paper relaxed the original restriction, using the particle swarm optimization and traversal method to determine the optimal parameters of the algorithm. This paper also combined the component filters of the original algorithm and picked the most suitable combination as the new collaborative algorithm.
    It was found that the original algorithm has already achieved the best possible outcome, apart from a few threshold parameters that need to be adjusted. The effective way to further enhance the performance is to attach only one NL-PCA filter before and after the pre-filtered rotationally invariant non-local mean (PRI-NLM) filter. Although the performance of the new collaborative algorithm is still a little short of advanced deep learning methods, it shows that the algorithm based on PCA denoising is indeed feasible. 
    It requires only a few parameters to be adjusted, and it is conceivable that they can be determined directly from the image, granting it a strong general capacity for various body parts, 
    and it merits further exploration.
    An auxiliary tool was also extracted from the new algorithm, encouraging further combination of it with other state-of-the-art methods to further improve their denoising performance.

\end{abstract}

    \section{Introduction}
        \hspace{1em}
        
    Magnetic resonance imaging can be used to view the three-dimensional structure and internal details of human tissues and organs, as well as non-invasive access to a large amount of physiological and pathological information in them, has a very important position in clinical diagnosis and scientific research \citep{ran_denoising_2019}, is currently one of the most widely used clinical medical imaging methods. However, there will always be noise during the acquisition process of MR images, which will not only reduce image quality and impact clinical diagnosis precision but will also interfere with later analysis tasks such as registration and segmentation. In some super-resolution algorithms, image denoising is required before the reconstruction step is performed\citep{coupe_collaborative_2013, manjon_non-local_2010}. Therefore, for further MR analysis or reconstruction, denoising algorithms that improve data quality during the imaging process are necessary. 

    Denoising algorithms are often inseparable from the estimation of the noise level, so the noise level estimation usually determines the performance of the denoising algorithm. One of the most commonly used methods in noise estimation is PCA (principle component analysis). Recently, many works in the literature have discussed and analyzed the relationship between noise variance and eigenvalues of the covariance matrix of patches in images under additive noise conditions \citep{zhao_detail-preserving_2018, chen_efficient_2015, fang_novel_2019, jiang_efficient_2019}. For noiseless images, most of their eigenvalues are close to 0, so patches in a clean image are usually located in a low-dimensional subspace. The deficiency of these works of literature is that the model is based on static Gaussian noise, while in reality, it is more common for the noise variance to change slowly with space. Noise in MRI obeys the Rician distribution, and local noise is usually underestimated, especially in low signal regions. Constant Rician noise obeys the Rayleigh distribution in the background of the MR image, so both local means and local variances of the background can be used for a robust estimation of the noise level\citep{aja-fernandez_noise_2008}. \cite{coupe2009object} proposed an object-based approach, which relies on the adjustment of the Median Absolute Deviation (MAD) estimator in the wavelet domain for Rician noise. In \cite{koay_analytically_2006}, the authors proposed an analytical method to solve the relationship between the variance of signal intensity in a magnitude MR image and the variance of Gaussian noise in two quadrature channels. When the mean and standard deviation of measured signal intensity are known, the underlying Gaussian noise can be computed pixel by pixel in the image. 
    Many studies in the field of estimation of the noise level of magnitude images have adjusted the "magnitude noise" to the actual noise level based on the findings of this paper \citep{coupe_robust_2010, manjon_adaptive_2010, liu_generalized_2014}. 

    Principal component analysis is often used for dimensionality reduction, but PCA can also be used for denoising. The key to PCA denoising is maximum separability, where the signal and noise are separated as sufficiently as possible by the orthogonal linear transformation. When conducting PCA decomposition on the image, it is observed that the parts related to the signal are predominantly concentrated in a limited number of components, while noise is evenly dispersed across all components. The eigenvectors corresponding to small eigenvalues are often associated with noise, and by discarding them, a certain degree of denoising is achieved. The adaptive version of PCA was first proposed by \cite{muresan_adaptive_2003}. This paper performed PCA decomposition on a localized group of image patches. Later, the algorithm was improved; that is, similar patches were first grouped before PCA decomposition was performed, and the process was continuously iterated to more effectively minimize noise\citep{zhang_two-stage_2010}. \cite{bao_structure-adaptive_2013} and \cite{lam_denoising_2014} proposed PCA-based denoising methods for DWI(diffusion weighted imaging).

   The NLM (non-local mean) method achieved excellent performance on the basis of the similarity between each two patches in the image. The core of the non-local mean method lies in weighted averaging, which takes advantage of the high degree of redundancy in patch similarity. NLM believes that all natural images contain excess information and that for each voxel in the image, there exist other voxels that are similar to it but may not be located in its spatial neighbor region. Several different versions of the NLM algorithms have been developed to date. In the course of development, these NLM algorithms not only preserve edges, but also enhance the signal-to-noise ratio and computational efficiency. The algorithm does not depend on the assumption of a specific image model. Early work is from \cite{buades_review_2005}, in which the idea of NLM and its corresponding algorithm were proposed for the first time.  NLM, originally used to denoise natural images, was developed further to denoise MR images \citep{manjon_non-local_2010}. The algorithm has been continuously improved, resulting in a reduced computational load \citep{coupe_optimized_2008}, as well as rotational invariance and integration with other denoising algorithms\citep{manjon_new_2012, manjon_mri_2015}. See \cite{bhujle_nlm_2019} for a comprehensive review of modern NLM algorithms.

    In \cite{bhujle_nlm_2019} the author compared several NLM algorithms for denoising Ricianly perturbed images and found that PRI-NLM3D \citep{manjon_new_2012} had the best performance in processing Ricianly perturbed MR images obtained from a single coil. The PRI-NL-PCA algorithm proposed by \cite{manjon_mri_2015} is similar to the former, but its guide images are obtained by a different pre-filter. The recently proposed deep learning methods, such as MCDnCNN\citep{jiang_denoising_2018} and RED-WGAN \citep{ran_denoising_2019} have better denoising performance than many traditional denoising methods, but require a large number of high-quality images for the training set. 
    This paper will broadly follow the idea of \cite{manjon_mri_2015} by using a PCA pre-filtered image to guide the weighted average in the NLM filter. However, this paper let the parameter used to define whether the eigenvalues are signals or noise to be determined, and this paper no longer decides to constrain the number of pixels in the 3D patches to be equal to the number of similar patches in each group. This paper found the optimal values of the relevant parameters using the particle swarm optimization and traverse method in order to achieve an improvement in the denoising performance of the original algorithm. 
    The experimental results demonstrate that the approach suggested by \cite{manjon_mri_2015} has achieved the optimum, with the exception of two or three threshold parameters that require adjustment. But it has been observed that using a median filter when the noise level is not significantly high will actually decrease the noise removal efficacy of the NL-PCA algorithm. In many combinations, it was found that the further improvement of the PRI-NL-PCA algorithm of \cite{manjon_mri_2015} can be maximized by connecting only one NL-PCA filter before and after the PRI-NLM filter while ensuring efficiency. 
    At the same time, this paper extracts an auxiliary combined tool from the algorithm that can further improve the performance of current state-of-the-art algorithms.
    
    To the best of our knowledge, our paper demonstrates for the first time the concept of complementary tools to further reduce noise in magnetic resonance images.

    The paper is organized as follows. The methods are described in section \ref{method}. The experiments and main results are given in section \ref{expr}. In section \ref{discussion}, the findings are discussed, and conclusions are made.
\section{Methods}
\label{method}
\hspace{1em}
    A denoising problem is usually defined as follows: Given a noisy image Y, treat it as the sum of the original noise-free image X and some noise n:
\begin{equation}
    Y=X+n
\end{equation}
    Thus, for any denoising algorithm, the goal is to find a good estimate $\hat{X}$ for a given noisy image. Researchers usually use metrics such as the Peak Signal Noise Ratio (PSNR) and the structural similarity index measure (SSIM) to evaluate how good an estimate is.
\subsection{Non-local PCA method}
\hspace{1em}
  
    This paper only looks for similar sets of patches in a search window. In a 3D window, for each 3D patch, this paper selects a set of (M-1) patches in the window that are the most similar to it and transforms those M patches into row vectors to form a sample matrix X. The values of the elements of the row vectors reflect the signal intensity. The similarity between two patches is measured by the Euclidean distance between their corresponding vectors. X is a matrix of $d^3\times M$, where d denotes the number of voxels contained in one edge of a square patch. The sample matrix is centralized to find its covariance matrix and further to find the covariance matrix's eigenvalues and eigenvectors. 
    
    Eigenvectors whose eigenvalues are less than $\tau\sigma$ are discarded, and $\sigma$ is calculated using the following equation:

    \begin{equation}
        \hat{\sigma}=\beta median\left( \sqrt{\lambda _s} \right) \,\,, \lambda _s=\left\{ \lambda _j|\sqrt{\lambda _j}\leqslant Tmedian\left( \sqrt{\lambda} \right) \right\} \,\,
        \label{1}
    \end{equation}
    where T is used to distinguish a set of eigenvalues as part of the signal and part of the noise. $\lambda _s$ is a short truncated version of the original eigenvalue array. The eigenvalues of the covariance matrix represent the magnitude of the variance of the data in the direction of the principal components.

    The remaining eigenvectors corresponding to the larger eigenvalues form a projection matrix $W^*$. $W^*\left( W^* \right) ^T$ is multiplied by the centralized sample matrix, and the result obtained after decentralization is the desired denoising matrix. Considering that each voxel is contained in more than one patch, the full estimate of the signal intensity and standard deviation for a voxel is combined by the uniform averaging rule. 

    \subsection{Prefiltered rotation invariant non-local means method} 
    \hspace{1em}
    For the traditional NLM estimator, the restored value of a certain voxel's signal intensity is the linear weighted average of the intensity of each voxel in a certain region of the image:
\begin{equation}
    \hat{v}\left( i \right) =\frac{\sum_{j\in \varOmega}{w\left( i,j \right) u\left( j \right)}}{\sum_{j\in \varOmega}{w\left( i,j \right)}}
    \label{11}
\end{equation}
    where v is the original signal intensity, u is the measured signal intensity, and $\Omega$ is the search volume around voxel i. Weight w(i, j) is allocated to u (j) when restoring voxel i.

    For patches with a structure similar to the reference patch, regardless of orientation, they should have a greater influence on it when averaging. Therefore, the original NLM filter needs to be made rotationally invariant. 

    If similar patches are searched from noisy images, the result may be susceptible to noise. The original image without noise can be used to more accurately measure the similarity between two patches. However, in practice, researchers often do not have such a perfect image, so it is needed to take a pre-denoised image as a guide instead. 
    \begin{equation}
w\left( i,j \right) =\begin{cases}
	\exp \left[ -\frac{1}{2}\left( \frac{\left( g\left( i \right) -g\left( j \right) \right) ^2+3\left( \mu _{Ni}-\mu _{Nj} \right) ^2}{2h_{i}^{2}} \right) \right] \,\,if\,\,\left| \mu _{Ni}-\mu _{Nj} \right|<h\\
	0 \hspace{16em} otherwise\\
\end{cases}
    \end{equation}
    The averages of the patches $N_i$ and $N_j$ that surround voxel i and j in the guide image are represented by $\mu _{Ni}$ and $\mu_{Nj}$. $h_i$ is correlated with the standard deviation of the noise in voxel i. The weights here are not yet normalized and should be calculated by multiplying them by the normalization factor to ensure that the sum of the weights corresponding to all voxels j in the search volume of voxel i is 1.

\subsection{Rician correction for image}
\label{Rcfi}
\hspace{1em}
In the presence of Rician noise, the first and second moments of the signal intensity can be expressed analytically by the equations
\begin{equation}
    \left< u \right> _{p_r}=\sqrt{\frac{\pi}{2}}\sigma _ge^{-\frac{v^2}{4\sigma _{g}^{2}}}\left[ \left( 1+\frac{v^2}{2\sigma _{g}^{2}} \right) I_0\left( \frac{v^2}{4\sigma _{g}^{2}} \right) +\frac{v^2}{2\sigma _{g}^{2}}I_1\left( \frac{v^2}{4\sigma _{g}^{2}} \right) \right] 
    \label{5}
\end{equation}
and
\begin{equation}
    \left< u^2 \right> _{p_r}=2\sigma _{g}^{2}+v^2
\end{equation}
where $\sigma _g$ is the Gaussian noise's standard deviation present in both the real and imaginary images, $p_r$ represents Rician distribution, $I_0$ and $I_1$ denote the zeroth and the first kind modified Bessel functions respectively. The meanings of u and v have been mentioned earlier.

    In the resulting image obtained through PCA denoising of the magnitude image, the signal intensity value for each voxel is determined by the uniform average of patches that contain the voxel. This method is similar to obtaining the first moment of the Rician distribution. However, according to \cite{koay_analytically_2006}, the first moment is biased and is not equivalent to the true signal intensity without noise. Thus, the formula (\ref{5}) and a pre-calculated look-up table must be utilized to transform the biased estimation into an unbiased one. 
    
 Meanwhile, utilizing the guide map to space-weight the original noisy MRI image by (\ref{11}), results in a biased estimate due to the Rician distribution. The second-moment formula of the Rician distribution will be used to construct the following unbiased estimator:
    \begin{equation}
        \hat{v}\left( i \right) =\sqrt{\max \left( \left( \frac{\sum_{j\in \varOmega}{w\left( i,j \right) u^2\left( j \right)}}{\sum_{j\in \varOmega}{w\left( i,j \right)}} \right) -2\sigma _{g}^{2}\left( i \right) \,\,, 0 \right)}
        \label{12}
    \end{equation}

\subsection{Noise estimation for Rician noise}

\hspace{1em}
    The estimate of noise affects not only 
    the Rician correction of the PCA-filtered image, but also the degree of smoothing of the non-local mean filter and the secondary Rician correction required during weighted averaging. Thus, an accurate noise estimation is crucial in optimizing the denoising performance.

\cite{manjon_mri_2015} obtained Gaussian-like local estimates using equation (\ref{1}), and then corrected them Ricianly. However, the noise estimation method mentioned above was not accurate enough at low noise level. 

The Rician noise in the area without NMR signal obeys the Rayleigh distribution, so \cite{aja-fernandez_noise_2008} used the distribution mode of the local mean of the sample in the background to calculate the Rician noise level. The following estimation method was adopted:
    \begin{equation}
        \sigma _g=\sqrt{\frac{2}{\pi}}median\left( \mu _{b} \right)
        \label{10}
    \end{equation}
    where $\mu_b$ denotes the local mean of the background. 
    Note that this estimation method is only applicable when the Rician noise level is constant in the image. 
    
    \cite{coupe2009object} proposed a MAD-based and object-based estimator designed for Rician noise. Note that the low sub-band LLL obtained from the 3D wavelet decomposition contains the feature information, while the highest sub-band HHH is mainly composed of coefficients corresponding to noise. The segmentation of the object is done in the LLL sub-band, and the mask obtained is used to extract coefficients $y_i$ corresponding to the object in the HHH sub-bands. When the noise follows the Gaussian distribution, its standard deviation $\sigma$ can be obtained by the following equation:
    \begin{equation}
        \hat{\sigma}=\frac{median\left( \left| y_i \right| \right)}{0.6745}
        \label{14}
    \end{equation}
    Under the condition of Rician noise, it is needed to perform additional analytical iterative correction to obtain unbiased estimates as \cite{koay_analytically_2006}.
For a detailed understanding of the object-based method mentioned above, this paper recommends referring to the original paper by \cite{coupe2009object}.

\subsection{The particle swarm optimization method}
\hspace{1em}
It is needed to find the set of parameters that make our NL-PCA filter optimal. This paper splits the original 3D image into closely spaced 3D windows of equal size, since searching the entire volume with highly overlapped windows will be time-consuming and not worth it when running the particle swarm optimization (PSO) algorithm in this work. Considering that the original 3D image is not necessarily divisible by the size of the 3D windows, after each dimension's final element, this paper adds padding to the corresponding array consisting of reflections mirrored from the edges by the padarray function in MATLAB. Then the effects of d, M, w, T, and $\tau\beta$ on algorithm performance are considered, where M is the number of similar patches in a group, w denotes the radius of a search window($\left( 2w+1 \right) ^3$ voxels lies in it), $\tau\beta$ and T are threshold parameters mentioned in section 2.1. 

The small eigenvalues of the covariance matrix of the sample points obtained by the PCA method are often related to noise, and the denoising can be realized to a certain extent by discarding their corresponding eigenvectors. A reasonable setting of $\tau\beta$ values can effectively remove noise components. To better define the noise part and the signal part, it is necessary to set the T value appropriately.  \cite{manjon_mri_2015} subjectively fixed T=2 in equation (\ref{1}) based on an example of PCA eigenvalues. In this article, T was a parameter that needed to be optimized. The parameter $\beta$ was used in \cite{manjon_mri_2015} to obtain the noise map, while in this study, it was multiplied with the parameter $\tau$ as a whole to remove small eigenvalues. Another weakness of \cite{manjon_mri_2015} in the analysis is that the constraint M =$d^3$ does not necessarily make the Peak Signal Noise Ratio (PSNR) reach the optimal value, and there is no theoretical basis to prove it.

For the PSO algorithm, this paper set SwarmSize to 50, maxIterations to 50, FunctionTolerance to 1e-3, and adopted the default values of MATLAB for the rest. 1$\%$ Gaussian noise was added to the ground truth and the PSO algorithm was run to obtain the set of optimal parameters (d, M, w, $\tau\beta$, T) when PSNR reached the maximum.

The deep learning method can achieve the same effect as the particle swarm optimization algorithm in this paper. However, it is initially believed that NL-PCA is a good traditional algorithm; that is, it should not rely too much on parameters. 

When the particle swarm optimization algorithm is used, to improve the search efficiency, this paper  makes the following qualification:
\begin{equation}
    \begin{cases}
	2\leqslant d\leqslant w+1\\
	d^3\leqslant M\leqslant \left( 2w+2-d \right) ^3\\
	2\leqslant w\leqslant 3\\
	\tau \beta \leqslant T
    \label{6}
    \end{cases}
\end{equation}
The meanings of d, M, $\tau\beta$, and T have been mentioned above. M= $(2w+2-d)^3$ corresponds to the situation in which all patches in the search window are selected to form a group. This paper originally let the upper limit of w be 4, but later found that this was not necessary. This is because it greatly increased the running time of the particle swarm optimization algorithm and did not necessarily help the denoising performance of the NL-PCA algorithm compete with those of the state-of-the-art methods. It also does not affect the final contributions of our paper.

When the number of samples is smaller than the data dimension in the sample matrix, the effect of PCA will be limited. This is caused by the singularity of the covariance matrix, which will cause the eigenvalues of part of the eigenvectors to become 0. According to the experiment, in this case, the smaller the number of samples, the more eigenvalues of the covariance matrix will be 0, the lower the threshold, and the image will hardly be denoised at last. For a better denoising effect, more samples need to be added to restore the covariance matrix to full rank. The above discussion explains the constraint $M\geqslant d^3$. 
$d\leqslant w+1$ because the left side of the inequality in the second line of (\ref{6}) must be smaller than the right side of that. $M\leqslant \left( 2w+2-d \right) ^3$ because the number of similar patches selected in a 3D window cannot exceed the total number of patches in the window.

On the other hand, according to the meanings of $\tau\beta$ and T mentioned earlier, the former should be less than the latter, otherwise the components associated with the signal will be discarded, leading to a decrease in denoising quality. 

\section{Experiments and results}
\label{expr}
\hspace{1em}
To facilitate the replicability of the experiments conducted, the MATLAB code utilized in our research will be accessible on the website: https://github.com/qizixinge/PCA-PRI-PCAr$\_$and$\_$PD.

\subsection{Experimental data description}
\hspace{1em}
BrainWeb MRI phantom \citep{collins_design_1998, kwan_mri_1999} was used in our experiment. The data set includes brain images of T1w, T2w and PDw, which can have a Rician noise level from $1\%$ to $15\%$ of the maximum signal intensity. The 3D image has 181*217*181 voxels. This paper selected images with $1mm^3$ voxel resolution for download. Only spatially isotropic noise distributions were used for the experiments.

The PSNR and the structural similarity index measure (SSIM) \citep{wang2004image} were used to measure the denoising performance of the algorithm:
\begin{equation}
    PSNR=20\log _{10}\frac{255}{RMSE}
\end{equation}
and
\begin{equation}
    SSIM\left( u,v \right) =\frac{\left( 2\mu _u\mu _v \right) \left( 2\sigma _{uv}+c_2 \right)}{\left( \mu _{u}^{2}+\mu _{v}^{2}+c_1 \right) \left( \sigma _{u}^{2}+\sigma _{v}^{2}+c_2 \right)}
\end{equation}
RMSE represents the root mean square error between the noiseless and filtered images. 
$c_1=\left( k_1L \right) ^2$ and $c_2=\left( k_2L \right) ^2$, where L denotes the range with a default value of 255, and $k_1=0.01 $, $k_2=0.03$. $\mu _u$ and $\mu _v$ are the local mean values of noisy image u and ground truth v, $\sigma _{uv}$ is the covariance of image u and v, $\sigma_u$ and $\sigma_v$ denote local standard deviations of u and v. The above six local estimates were completed by each voxel in the $3\times 3\times 3$ neighborhood. The global SSIM value is calculated by averaging all local SSIM estimates. Compared to PSNR, SSIM is more in line with our visual characteristics. For clarity, the values of PSNR and SSIM are only evaluated within the region of interest (ROI), i.e. the head tissues, which are obtained by excluding the background region with a signal intensity of 0 in the ground truth.

\subsection{The optimal parameters}
\hspace{1em}
The approximate optimal results for the NL-PCA filter by PSO are d = 3, M = 27, w = 3, $\tau\beta$ = 2.46, T = 2.46. In fact, there was an order of magnitude difference of $10^{-2}$ between the optimal values of $\tau\beta$ and T obtained by our initial PSO algorithm. This paper fixed (d, M, w) and let $\tau\beta$ and T vary, and finally found that $\tau\beta$ = T when the optimal denoising performance was achieved for the NL-PCA algorithm. This result meets our expectations and is consistent with the product of $\tau$=2.1 and $\beta$=1.16 given in \cite{manjon_mri_2015} (where T=$\infty $). 

\cite{manjon_mri_2015} obtained the optimal parameter of the NL-PCA filter (d, M, w) = (4,64,3) with the limitation M=$d^3$, which exactly satisfies M= $(2w+2-d)^3$. Taking into account the mathematical principle of the PCA method, the eigenvalue of the covariance matrix only needed to be calculated once in each search window, which will greatly improve the efficiency of the code. However, when (d, M, w) = (3,27,3), for each patch in a window, it is necessary to search for the 26 most similar patches in the other 124 patches. If the number of overlapping voxels between windows is increased, the code running time will grow as a power function. Although (d, M, w) = (4,64,3) is not the optimal solution, the denoising performance corresponding to it can be greatly improved by overlapping windows and even exceeds that corresponding to our optimal solution. To improve the efficiency of the NL-PCA filter when (d, M, w) = (3,27,3), this research only looks for patches in each window that are most similar to its central patch. This paper lets step=2 between consecutive windows. But the images obtained in the experiment were even blurrier than the initial noisy images. According to the above phenomena and the mathematical characteristics of the PCA method, it is believed that the key point of the NL-PCA filter is not non-local, but the selection of all patches in search windows.

We can also infer from the experimental results that when $\frac{M}{d^3}$=1, the best performance of the NL-PCA filter can be obtained by satisfying $\tau\beta$=T=2.46, regardless of the parameter w and the overlap between search windows.

We also tried other combinations, such as (d, M, w)=(3, 125, 3), with the corresponding step=2, and obtained the optimal solution $\tau\beta$=T$\approx$1.5. Our result is that the PSNR of the denoised image at various noise levels is slightly lower compared to the case where (d, M, w, $\tau\beta$, T)=(4, 64, 3, 2.46, 2.46).

Finally, this paper was able to determine the optimal parameter set, and let step=3 between windows in the NL-PCA process to improve filter efficiency.

\subsection{Comparison}
\subsubsection{The necessity of median filter}
\hspace{1em}
\begin{table}[ht]
\centering
\caption{PSNR and SSIM measurement of different methods on T1w image with different noise levels. The first row of data in the table below is the PSNR and the second row is the SSIM. The following tables are similar.}
\begin{tabular}{llllll}
\hline
\multirow{2}{*}{Denoising type}        & \multicolumn{5}{l}{noise level}                 \\ \cline{2-6} 
                                       & 1\%     & 7\%     & 13\%    & 19\%    & 25\%    \\ \hline
\multirow{2}{*}{Without median filter} & 43.8716 & 27.4234 & 21.4615 & 17.5877 & 14.6929 \\
                                       & 0.9591  & 0.7740  & 0.6668  & 0.5703  & 0.4838  \\
\multirow{2}{*}{With a median filter}    & 34.3526 & 26.8655 & 21.6030 & 17.9562 & 15.1761 \\
                                       & 0.9272  & 0.7407  & 0.6266  & 0.5244  & 0.4341  \\ \hline
\end{tabular}
\label{b3}
\end{table}
As mentioned by \cite{manjon_mri_2015}, the median filter can significantly promote patch selection in the grouping process of NL-PCA denoising, but our experimental results (see Table \ref{b3}) show that applying a median filter under low-level noise will have a negative effect on the denoising process. The median filter does enhance the homogeneity of patches within the group, providing a sparser representation, but at the same time, the large eigenvalues representing the signal are also reduced. The median filter will make noisy images noisier. The noise variance will be even larger than the local variance of the signal in the image under large-level noise, and the large eigenvalues of the covariance matrix correspond to the noise, which can be effectively reduced by the median filter to achieve the effect of noise reduction. The higher the noise level, the smaller the relative impact on the eigenvalues corresponding to the signal variance. Therefore, it can be seen that the PSNR of the experimental group with the median filter is higher at a high noise level. But after all, the original signal variance has been affected, so the experimental group without the median filter always has a higher SSIM than the group with the median filter. 
Based on the experimental results, this research will no longer use the median filter in the NL-PCA algorithm.

\subsubsection{Analysis of denoising effect under different combinations}
\label{3_3_2}
\hspace{1em}
In this section, the author compares the denoising performance of the algorithm under the same filter iteration and different filter combinations.  
Five experiments were first conducted: only the NL-PCA method was used to denoise images disturbed by Rician noise; the NL-PCA filter was used twice; the noisy images processed by the NL-PCA method were corrected Ricianly; the corrected images were filtered by the NL-PCA method for the second time; the Rician corrected image was applied to a PRI-NLM filter. The above five groups of denoising types are referred to as d, dd, dg, dgd, and dgp respectively. The experimental results are shown in Table \ref{b2}, where n means the original noisy image, d represents the denoised image obtained by the NL-PCA method, g represents the guide image from Rician correction, and p represents the final image acquired by denoising with the PRI-NLM filter. To minimize the influence of noise level estimation on the results of the simulation experiment, the formula (\ref{10}) was adopted to estimate the noise level for g and p. 
\begin{table}[]
\centering
\caption{PSNR and SSIM measurement of different combinations on T1w normal image with different noise levels.}
\begin{tabular}{llllll}
\hline
\multirow{2}{*}{\begin{tabular}[c]{@{}l@{}}Denoising types of\\  T1w normal images\end{tabular}} & \multicolumn{5}{l}{noise level}                                                              \\ \cline{2-6} 
                                                                                                 & 1\%              & 3\%              & 5\%              & 7\%              & 9\%              \\ \hline
\multirow{2}{*}{n} & 39.9282 & 30.0969 & 25.5535 & 22.5785 & 20.3697 \\
& 0.9344 & 0.7217 & 0.5594 & 0.4413 & 0.3548 \\
\multirow{2}{*}{d}                                                                               & 43.8672          & 35.0191          & 30.5097          & 27.4302          & 25.0514          \\
                                                                                                 & 0.9591           & 0.8679           & 0.8158           & 0.7741           & 0.7367           \\
\multirow{2}{*}{dd}                                                                              & 43.8283          & 35.0494          & 30.5359          & 27.4523          & 25.0706          \\
                                                                                                 & 0.9592           & 0.8686           & 0.8172           & 0.7763           & 0.7395           \\
\multirow{2}{*}{dg}                                                                              & 44.8635          & 38.4251          & 35.3376          & 33.1685          & 31.4278          \\
                                                                                                 & 0.9897           & 0.9515           & 0.9085           & 0.8657           & 0.8229           \\
\multirow{2}{*}{dgd}                                                                             & 44.8489          & 38.5927          & 35.5587          & 33.4259          & 31.7133          \\
                                                                                                 & 0.9900           & 0.9544           & 0.9141           & 0.8741           & 0.8336           \\
\multirow{2}{*}{dgp}                                                                             & 44.8978          & 39.0869          & 36.1373          & 34.0693          & 32.4144          \\
                                                                                                 & 0.9900           & 0.9576           & 0.9262           & 0.8956           & 0.8637           \\
\multirow{2}{*}{dgpp}                                                                            & 44.3191 & 38.5981 & 35.6442 & 33.6161 & 32.0543 \\
                                                                                                 & 0.9890  & 0.9542  & 0.9229  & 0.8939  & 0.8644  \\
\multirow{2}{*}{dgpd}                                                                            & \textbf{45.3965}          & \textbf{39.4307}          & \textbf{36.4272}          & \textbf{34.3510}          & \textbf{32.7010}          \\
                                                                                                 & \textbf{0.9908}           & \textbf{0.9607}           & \textbf{0.9313}           & \textbf{0.9028}           & \textbf{0.8728}           \\

\hline
\end{tabular}
\label{b2}
\end{table}

Taking into account all noise levels comprehensively, the denoising performance can be obtained: $d<dd<dg<dgd<dgp$.  
Inspired by the above five groups of experiments, this paper tried to iteratively use the PRI-NLM filter to conduct the sixth group of experiments.  This research used the final denoised image obtained by the dgp group to guide the PRI-NLM filter to denoise the original noisy image again. The obtained denoised image as the guide image of the next filtering may constantly approximate the ground truth and thus continuously improve the PSNR of the image. However, experimental results show that when the PRI-NLM method is used for the same noisy image iteratively, the denoising performance will degrade, no matter how accurate the noise level estimation is.

On the other hand, this paper considered using the NL-PCA method to filter the denoised images obtained from the dgp group adaptively and performed the seventh group of experiments. The denoised images obtained by the PRI-NLM algorithm retained a small amount of unknown distributed noise. NL-PCA, as a good adaptive filter, can effectively solve the problem. The discussion above was confirmed by the experimental results of the dgpd group. 

The experimental results for T2w MS lesion images are shown in Table \ref{b4}. This article found that the denoising performance of the dgpd group was not optimal at low noise level. On the one hand, in the NL-PCA process, the majority of signal-related parts are concentrated in a few specific components, while noise is distributed evenly across all components. The eigenvectors associated with small eigenvalues are typically indicative of noise. On the other hand, when the NL-PCA method achieves optimal denoising performance for a certain type of image, it should satisfy $\tau\beta$=T. If the value of T is too small, there is still much noise residue; if T is too large, large signal components will also be discarded. During the experiment, it was found that for T2w images, when the NL-PCA filter's (d, M, w) was set to (4, 64, 3) unchanged, T and $\tau\beta$ should be set to smaller values to improve denoising performance. This indicates that the noisy image under the optimal parameters of the T1w images simultaneously loses some of the original signal during the NL-PCA filter denoising process. This well explains the phenomenon that there is no significant improvement in the dgpd groups compared to the dgp group at 1$\%$ level of noise. For different types of images, adjusting the parameters of the NL-PCA filter in a reasonable way will result in results similar to Table \ref{b2}, where the dgpd group performs optimally at all noise levels. Overall, according to Table \ref{b2} and \ref{b4}, for different types of images, adding an adaptive filter after the PRI-NLM method can improve the denoising performance of the original algorithm to a certain extent.

\begin{table}[]
\centering
\caption{PSNR and SSIM measurement of different combinations on T2w MS lesion image with different noise levels.}
\begin{tabular}{llllll}
\hline
\multirow{2}{*}{\begin{tabular}[c]{@{}l@{}}Denoising types of\\ T2w MS lesion images\end{tabular}} & \multicolumn{5}{l}{noise level}                                                              \\ \cline{2-6} 
                                                                                                   & 1\%              & 3\%              & 5\%              & 7\%              & 9\%              \\ \hline
\multirow{2}{*}{n} & 39.8933 & 29.9163 & 25.2907 & 22.2712 & 20.0251 \\
& 0.9194 & 0.6959 & 0.5556 & 0.4629 & 0.3962\\
\multirow{2}{*}{d}                                                                                 & 42.4305          & 33.4971          & 29.0037          & 26.0049          & 23.7453          \\
                                                                                                   & 0.9433           & 0.8155           & 0.7515           & 0.7095           & 0.6768           \\
\multirow{2}{*}{dd}                                                                                & 42.0224          & 33.4951          & 29.0208          & 26.0231          & 23.7629          \\
                                                                                                   & 0.9433           & 0.8161           & 0.7528           & 0.7113           & 0.6793           \\
\multirow{2}{*}{dg}                                                                                & 43.4931          & 37.0679          & 33.8937          & 31.7608          & 30.0978          \\
                                                                                                   & 0.9881           & 0.9412           & 0.8856           & 0.8348           & 0.7891           \\
\multirow{2}{*}{dgd}                                                                               & 43.0035          & 37.1544          & 34.0829          & 31.9980          & 30.3754          \\
                                                                                                   & 0.9883           & 0.9450           & 0.8935           & 0.8460           & 0.8030           \\
\multirow{2}{*}{dgp}                                                                               & \textbf{43.6912}          & 37.2501          & 34.3313          & 32.3436          & 30.7588          \\
                                                                                                   & 0.9879           & 0.9462           & 0.9012           & 0.8619           & 0.8269           \\
\multirow{2}{*}{dgpp}                                                                              & 43.3181 & 36.7526 & 33.9046 & 31.9675 & 30.3988 \\
                                                                                                   & 0.9867  & 0.9426  & 0.8976  & 0.8604  & 0.8283  \\
\multirow{2}{*}{dgpd}                                                                              & 43.5902          & \textbf{37.5880}          &\textbf{34.6575}          & \textbf{32.6646}          & \textbf{31.0859}          \\
                                                                                                   & \textbf{0.9885}           & \textbf{0.9502}           & \textbf{0.9088}          & \textbf{0.8722}           & \textbf{0.8398}           \\

\hline
\end{tabular}
\label{b4}
\end{table}

\subsection{Theoretical limits of PRI-NLM filters}
\hspace{1em}
Surprisingly, the PRI-NLM filter usually can produce a denoised image with a higher PSNR than the image that guides it. If the original ground truth image and the most accurate noise map were directly input into the filter, it would reach the results shown in Table \ref{bend}.


\begin{table}[]
\centering
\caption{PSNR and SSIM measurement of theoretical limits on T1w image with different noise levels}
\label{bend}
\begin{tabular}{llllll}
\hline
\multirow{2}{*}{\begin{tabular}[c]{@{}l@{}}Denoising types of\\ T1w normal images\end{tabular}} & \multicolumn{5}{l}{noise level}                 \\ \cline{2-6} 
                                                                                                & 1\%     & 3\%     & 5\%     & 7\%     & 9\%     \\ \hline
\multirow{2}{*}{p}                                                                              & 48.6765 & 43.6352 & 40.6080 & 38.6111 & 37.1053 \\
                                                                                                & 0.9965  & 0.9812  & 0.9623  & 0.9469  & 0.9313  \\
\hline
\end{tabular}
\end{table}

This paper set the h parameter unchanged. The results in Table \ref{bend} indirectly indicate that PRI-NLM will play a negative role when the guide image is close enough to the original noiseless image. Even if the noiseless original image is used as a guide image, the noisy image cannot be denoised completely. However, they also show that the denoising potential of the PRI-NLM filter can even exceed that of current mainstream deep learning methods such as MCDnCNN \citep{jiang_denoising_2018} and RED-WGAN \citep{ran_denoising_2019}, and the denoising performance of this traditional filter is fully explored.

This paper would next explore the mathematical principles that lead to this experimental phenomenon. For 3D MRI images, the relationship between the measured signal intensity and the original signal intensity is as follows:

\begin{equation}
u^2=\left( v+n_1 \right) ^2+n_{2}^{2}
\end{equation}

$n_1$ and $n_2$ are independent of each other and both satisfy the Gaussian distribution with standard deviation $\sigma_g$. According to (\ref{12}), the estimated voxel signal strength obtained by using the PRI-NLM filter satisfies

\begin{align}
\hat{v}^2\left( i \right) &=\sum_{j\in \varOmega}{\tilde{w}\left( i, j \right) u^2\left( j \right)}-2\sigma _{g}^{2}\notag
\\&=\sum_{k=1}^{k^*}{\sum_{j\in I_k}{\tilde{w}\left( i, j \right) u^2\left( j \right)}}-2\sigma _{g}^{2}\notag
\\&=\varphi +\sum_{k=1}^{k^*}{v^2\left( i_k \right) \left( \sum_{j\in I_k}{\tilde{w}\left( i, j \right)} \right)}\notag
\\&=\varphi +\sum_{k=1}^{k^*}{v^2\left( i_k \right) \widetilde{W}\left( i, i_k \right)}
\label{13}
\end{align}

where $\tilde{w}$ is the normalized weight, and $\tilde{W}$ is the sum weight corresponding to $v^2\left( i_k \right) $. Equation (\ref{13}) consists of two parts: disturbance term and mean squared signal term. It can be seen that the filter will also affect the original signal during the process of noise reduction. This paper divides the search volume into k* disjoint subsets $I_1$, $I_2$, ..., $I_k$, ..., $I_{k^*}$. The original voxel signal strength in each subset is the same as the other. Select one voxel $i_k$ from each of the above subsets and let $i_{k^*}$=i. $\varphi$ in (\ref{13}) satisfies the following formula:

\begin{equation}
\varphi =2\sum_{k=1}^{k^*}{v\left( i_k \right) \sum_{j\in I_k}{n_{j1}\tilde{w}\left( i, j \right)}}+\sum_{j\in \varOmega}{\tilde{w}\left( i, j \right) \left( n_{j1}^{2}+n_{j2}^{2} \right)}-2\sigma _{g}^{2}
\end{equation}
The ensemble average of $\varphi$ is 0. At this point, it is clear that $\hat{v}^2\left( i \right) \ne v^2\left( i \right) $, even if the author let the $\varphi$ be 0. This is why even when the noise-free image is directly used to guide the PRI-NLM filter, the noisy image cannot be completely restored.

\subsection{Comparison of the performance of several auxiliary denoising tools}
\hspace{1em}
The PRI-NLM filter is a powerful tool to improve the denoising quality of the initial algorithm. Based on the results of Table \ref{b2} and \ref{b4}, its performance will be further enhanced by combining it with other adaptive filters.

The purpose of this section is to compare several filter combinations and select the best tool that can further improve the denoising quality even for the current advanced algorithms. A typical representative of the state-of-the-art Rician noise removal method is MCDnCNN \citep{jiang_denoising_2018}. 
For simplicity, it was assumed that the residual noise obeys the Gaussian distribution. That is, this paper instead built low-noise 3D images. This paper directly adopted the PSNR measures of the noise-specific model MCDnCNN method in \cite{jiang_denoising_2018} on the Brainweb dataset and took the corresponding RMSE directly as the standard deviation of Gaussian noise. Web Plot Digitizer was used to extract data from the original literature line chart. The final calculation results are shown in Table \ref{b5} where c stands for the constructed alternative image. p and d have the same meanings as before. The noise map required by the PRI-NLM filter had been precisely provided by (\ref{10}).  As can be seen in the tables, the denoising performance $cpd>cp>cpdp>cpp$.

\begin{table}[htbp]
\caption{ PSNR and SSIM measurement of different combinations on the simulated prefiltered T1w normal image with different noise levels}
\centering
\begin{tabular}{llllll}
\hline
\multirow{2}{*}{\begin{tabular}[c]{@{}l@{}}Denoising types of\\ T1w normal images\end{tabular}} & \multicolumn{5}{l}{noise level}                                                              \\ \cline{2-6} 
                                                                                                & 1\%              & 3\%              & 5\%              & 7\%              & 9\%              \\ \hline
\multirow{2}{*}{c}                                                                              & 44.9995          & 39.6092          & 36.7519          & 34.9294          & 33.4301          \\
                                                                                                & 0.9856           & 0.9533           & 0.9164           & 0.8819           & 0.8457           \\
\multirow{2}{*}{cp}                                                                             & 46.1253          & 41.5109          & 39.0865          & 37.4210          & 36.1162          \\
                                                                                                & 0.9936           & 0.9766           & 0.9579           & 0.9406           & 0.9242           \\
\multirow{2}{*}{cpd}                                                                            & \textbf{47.3405} & \textbf{42.3020} & \textbf{39.5792} & \textbf{37.7697} & \textbf{36.3932} \\
                                                                                                & \textbf{0.9949}  & \textbf{0.9788}  & \textbf{0.9604}  & \textbf{0.9434}  & \textbf{0.9273}  \\
\multirow{2}{*}{cpp}                                                                            & 45.1147          & 39.7345          & 36.9483          & 35.1027          & 33.7467          \\
                                                                                                & 0.9921           & 0.9694           & 0.9440           & 0.9205           & 0.8982           \\
\multirow{2}{*}{cpdp}                                                                           & 45.9375          & 40.1223          & 37.0625          & 35.1204          & 33.7387          \\
                                                                                                & 0.9931           & 0.9705           & 0.9443           & 0.9204           & 0.8980           \\ \hline
\end{tabular}
\label{b5}
\end{table}

From Table \ref{b5}, it can be inferred that even for the state-of-the-art algorithms, the PRI-NLM method can still achieve an improvement in their performance. The premise is that the algorithm used for pre-filtering can remove noise without losing too much original signal, and provide a good guide image. 

Each use of the PRI-NLM filter only changes the normalized weights in the formula (\ref{13}). In the process of running the code, it was found that multiple applications of the PRI-NLM filter will cause the weight $\widetilde{W}\left( i, i_{k^{*}} \right) $ of voxels in the search area to become smaller on the whole, that is, the deviation from the ground truth will become larger and larger. The above discussion can be further supported by the experimental results in Table \ref{b2} and \ref{b4}.

An adaptive filter can effectively suppress the noise corresponding to $\varphi$ in (\ref{13}), so the noise removal performance of the cpd group is better than that of the cp group. However, if the image obtained by the cpd group continues to be used as a guide image to apply a PRI-NLM filter to the original noisy image, it still cannot avoid the overall decrease of $\widetilde{W}\left( i, i_{k^{*}} \right) $, resulting in a decrease in noise reduction quality.

From the combination of pdp, we naturally think of the combination of dpd. However, it should be noted that the NL-PCA filter is particularly good at handling Gaussian noise in images under the original optimal parameter conditions, so the constructed alternative image cannot be directly used to determine the performance of the auxiliary denoising combination of dpd. Instead, this paper will use the images from the dg group in section \ref{3_3_2} for analysis. In theory, the auxiliary tool corresponding to dpd can better improve the performance of the original algorithm than pd, because the guide images of the previous set used in the PRI-NLM process deviate less from the ground truth. However, our experiment found that the improvement of dpd compared to pd is extremely small or even negligible.

\begin{figure}[htbp]
  \centering
  \begin{minipage}[b]{0.4\linewidth}
    \centering
    \includegraphics[width=\textwidth]{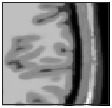}
    (a)
  \end{minipage}%
  \begin{minipage}[b]{0.4\linewidth}
    \centering
    \includegraphics[width=\textwidth]{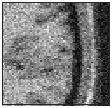}
    (b)
  \end{minipage}\\
  \begin{minipage}[b]{0.4\linewidth}
    \centering
    \includegraphics[width=\textwidth]{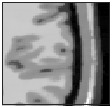}
    (c)
  \end{minipage}%
  \begin{minipage}[b]{0.4\linewidth}
    \centering
    \includegraphics[width=\textwidth]{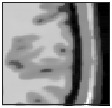}
    (d)
  \end{minipage}\\
\begin{minipage}[b]{0.4\linewidth}
    \centering
    \includegraphics[width=\textwidth]{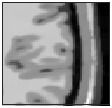}
    (e)
  \end{minipage}%
  \begin{minipage}[b]{0.4\linewidth}
    \centering
    \includegraphics[width=\textwidth]{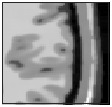}
    (f)
  \end{minipage}\\
  \caption{One denoised T1w normal example with a noise-free image(a), the noisy image(b), denoised image of cp(c), cpp(d), cpd(e) and cpdp(f) from the Brainweb dataset. A magnified view is presented to better distinguish the distinctions between the compared groups.}
  \label{t4}
\end{figure}

\begin{figure}[htbp]
  \centering
  \includegraphics[width=0.48\linewidth]{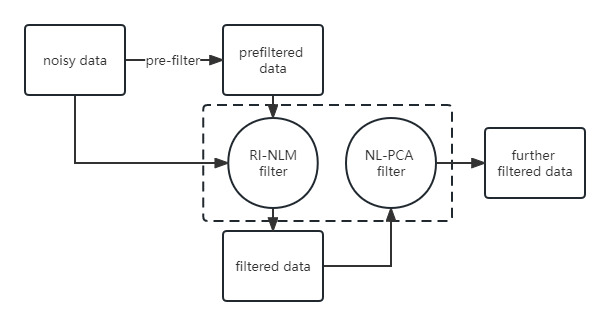}
  \caption{The framework of our proposed scheme. The interior of the dashed box displays the structure of PD.}
  \label{t7}
\end{figure}

Figure \ref{t4} shows the visual differences between the different groups of results with 9$\%$ Rician noise. From a subjective point of view, the denoising effect of the cp and the cpd group is not much different from that of the cpp and the cpdp group visually, but it can be observed in some sulcus areas that some fine details are excessively smoothed in the latter two groups. We will add an NL-PCA filter after the PRI-NLM filter and this combination will be called PD for short. The structure of our proposed PD is shown in Figure \ref{t7}.

\subsection{Appraisal on clinical data}
\hspace{1em}
Simulation data which are simplifications of real images may not accurately represent realistic details. This paper used two datasets to qualitatively test the consistency of the performance improvement using PD on clinical data. The auxiliary combination tool still uses images after NL-PCA filtering and Rician correction as the target for improvement. This paper uses the object-based method \citep{coupe2009object} based on (\ref{14}) to estimate the Rician noise level of clinical data.

\begin{figure}[htbp]
  \centering
  \begin{minipage}[b]{0.3\linewidth}
    \centering
    \includegraphics[width=\textwidth]{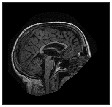}
  \end{minipage}%
  \begin{minipage}[b]{0.3\linewidth}
    \centering
    \includegraphics[width=\textwidth]{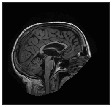}
  \end{minipage}%
  \begin{minipage}[b]{0.3\linewidth}
    \centering
    \includegraphics[width=\textwidth]{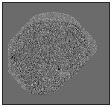}
  \end{minipage}\\
\begin{minipage}[b]{0.3\linewidth}
    \centering
    \includegraphics[width=\textwidth]{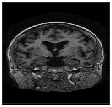}
  \end{minipage}%
\begin{minipage}[b]{0.3\linewidth}
    \centering
    \includegraphics[width=\textwidth]{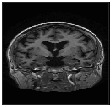}
  \end{minipage}%
  \begin{minipage}[b]{0.3\linewidth}
    \centering
    \includegraphics[width=\textwidth]{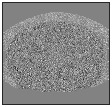}
  \end{minipage}\\
    \begin{minipage}[b]{0.3\linewidth}
    \centering
    \includegraphics[width=\textwidth]{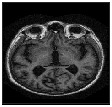}
  \end{minipage}%
  \begin{minipage}[b]{0.3\linewidth}
    \centering
    \includegraphics[width=\textwidth]{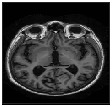}
  \end{minipage}%
  \begin{minipage}[b]{0.3\linewidth}
    \centering
    \includegraphics[width=\textwidth]{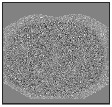}
  \end{minipage}
  \caption{Example results of the improved NL-PCA method on real data from the OASIS dataset. From left to right: Original noisy image, filtered image, and the corresponding residuals.}
  \label{t5}
\end{figure}

The first was a magnetization prepared rapid gradient-echo (MP-RAGE) T1w volumetric sequence from the Open Access Series of Imaging Studies (OASIS) database acquired on a  1.5-T Vision scanner(Siemens, Erlangen, Germany) with TR=9.7ms, TE=4.0ms, TI=20ms, TD=200ms, flip angle=10°, voxel resolution=1×1×1.25$mm^3$,  and 256×256×128 voxels\citep{marcus2007open}. The Rician noise level was estimated to be 2.36$\%$ of the maximum intensity. Figure \ref{t5} shows the example results of filtering this dataset. No significant anatomical information exists in any residual images, indicating that the proposed method successfully removes noise. This dataset took 295 seconds to process with the NL-PCA filter and 186 seconds with the PRI-NLM filter.

\begin{figure}[htbp]
    \centering
    \begin{minipage}{0.5\textwidth}
        \centering
        \includegraphics[width=\textwidth]{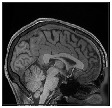}
        (a)
    \end{minipage}%
    \begin{minipage}{0.5\textwidth}
        \centering
        \begin{minipage}{0.45\linewidth}
            \includegraphics[width=\linewidth]{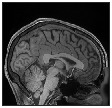}
            \centering
            (b)
        \end{minipage}%
        \begin{minipage}{0.45\linewidth}
            \includegraphics[width=\linewidth]{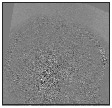}
            \centering
            (c)
        \end{minipage}\\
        \begin{minipage}{0.45\linewidth}
            \includegraphics[width=\linewidth]{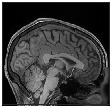}
            \centering
            (d)
        \end{minipage}%
        \begin{minipage}{0.45\linewidth}
            \includegraphics[width=\linewidth]{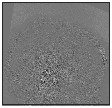}
            \centering
            (e)
        \end{minipage}
    \end{minipage}
    \caption{One denoised T1w example from the HCP dataset with (a)original noisy image, (b)denoised image with auxiliary tools corresponding to pd, (c)residual of (b), (d)denoised image without the auxiliary tools, (e)residual of (d). }
    \label{t6}
\end{figure}

The second was a 3D MP-RAGE T1w volumetric sequence from the Human Connectome Project (HCP) database \citep{GLASSER2013105, JENKINSON2012782, FISCHL2012774, JENKINSON2002825} acquired on a customized Siemens Skyra 3T scanner with TR=2400ms, TE=2.14ms, TI=1000ms, flip angle=8°, voxel resolution=0.7×0.7×0.7$mm^3$, and 260×311×260 voxels\citep{elam2021human}. The Rician noise level was estimated to be 0.28$\%$ of maximum intensity. Figure \ref{t6} provides a comparison of filtering results corresponding to images of type dgpd and dg on this dataset. It was noticed that both of these schemes performed very well on this dataset, but from the background of residual images, it can be seen that the former performs better in removing noise compared to the latter. Excellent denoising performance will be beneficial for some fine anatomical details of the image to become clearer. This dataset took 901 seconds to process with the NL-PCA filter and 517 seconds with the PRI-NLM filter.

\section{Disscussion}
\label{discussion}
\hspace{1em}
Currently, there are not many pieces of research that integrate the PRI-NLM filter with other traditional filters or deep learning methods, let alone PD. In \cite{manjon2019mri}, the proposed method based on the 3D patch-based convolutional neural network combined with a RI-NLM filter showed a performance comparable to that of the PRI-NLPCA method under low and medium noise conditions, and surpassed the latter under high noise levels. 

In this paper, the author first revised the trimmed median method in \cite{manjon_mri_2015} in an attempt to overcome its subjective shortcomings. This article also tried to relax the restriction of M=$d^3$ and explore the parameter sets (d, M, w, $\tau\beta$, T) that can optimize the performance of the NL-PCA filter. This research finds the optimal solution using the particle swarm optimization and traversal method. However, although the NL-PCA method has an internal noise estimation technique, the results obtained by the NL-PCA method are not accurate enough at low noise level, which is greatly disturbed by the original signal. This paper used the control variable method, providing a true noise map for later Rician correction and the PRI-NLM filter. The experimental results verify that the scheme proposed by \cite{manjon_mri_2015} is optimal. The effective way to improve its performance is to attach a single NL-PCA filter both in front and behind the PRI-NLM filter. Contrary to the original literature, the use of a median filter when the noise level is not particularly high will reduce the noise reduction effect of the NL-PCA algorithm. This paper analyzes the theoretical limit of the PRI-NLM filter and finds that the combined auxiliary tool PD obtained by following it with only one NL-PCA filter improved the performance of the initial algorithm better than other combinations.

\cite{manjon_mri_2015} argued that the NL-PCA filter benefited from sparseness and obtained a very sparse representation by grouping similar patches together. However, it was found in the experiment that the core of the filter is to select all patches in search windows for PCA decomposition. This paper has proposed two schemes based on sparsity. One is to select other patches similar to each patch in a window to form a group. The second is to only select patches similar to the central patch in each window. The first PCA scheme, which has good noise reduction performance, but is still inferior to \cite{manjon_mri_2015}, may be caused by the complete lack of overlap between search windows; the second one speeds up the code but makes the image more blurry than the original noisy image.

PD is an auxiliary tool to improve the performance of NL-PCA pretreatment and high PSNR simulation predenoising. In general, the benefit of the noise reduction of the PRI-NLM filter is greater than the loss of signal mean square, and the NL-PCA filter further suppressed the disturbance term in the unbiased estimation on this basis. Only one NL-PCA filter is needed in PD, and the improvement brought about by increasing its number is negligible. This paper constructed guide images with low Gaussian noise levels to replace the denoising results of the state-of-the-art methods. The actual improvement brought about by PD may be lower than the simulation results obtained in the experiment. But at least the latter provides theoretical feasibility for improving the denoising performance by combining the auxiliary tool the author proposed with other state-of-the-art algorithms, laying a foundation for future related research.  Of course, there are exceptions, such as when working with OASIS clinical data, it was found that additional use of PD in the NL-PCA filter did not show significant changes compared to not using it. It should be noted that the PRI-NLM filter contained in PD only applies to isotropic voxels, while the voxels of the images in the OASIS dataset are anisotropic.

This paper verifies the optimality of the NL-PCA scheme in \cite{manjon_mri_2015} and improves the objectivity of the parameter selection of the algorithm.\cite{manjon_mri_2015} fixed T=2 with an example of PCA eigenvalues only, and this paper obtained the optimal solution of $\tau\beta$ and T through calculation and theoretical analysis. At the same time, the PD tool was also extracted, and the internal NL-PCA filter can be replaced with other stronger adaptive filters. A good prefiltered image and noise map are required for PD,  and the former can be obtained from the Rician correction of the image after being denoised by an adaptive filter or from deep learning methods. Of course, a PD tool and a NL-PCA filter can be synthesized to obtain a stronger algorithm than PRI-NL-PCA, we call it the PCA-PRI-PCAr algorithm, where r represents the Rician correction. The application of the PD tool and the PCA-PRI-PCAr algorithm proposed in this paper can improve the quality of MRI data for further registration, segmentation, or reconstruction. In clinical diagnosis, they can improve the visibility of intricate anatomical details.

For the parameter group (d, M, w, $\tau\beta$, T) of the NL-PCA filter, the optimal solution of $\tau\beta$ and T is (2.46, 2.46). Compared with (2.2×1.29, 2) in \cite{manjon_mri_2015}, our NL-PCA filter only slightly improved under low level noise. The results show that the NL-PCA filter has similar performance for different types of images under the two sets of parameters, but the latter can estimate the noise level more accurately. In the future, we can study to find the optimal parameters $\tau\beta$ and T that can obtain a sufficient accurate noise estimation under the condition that the noise removal performance of the NL-PCA filter is not greatly affected. For images with different types and even different tissues, $\tau\beta$ and T which enable the NL-PCA filter to achieve the optimal performance or obtain the most accurate noise estimation are generally different. In the future, algorithms can be designed to automatically calculate the optimal T and $\tau\beta$ from the image itself. To keep things simple, we can look at the case $\tau\beta$=T. We can also change the way of thinking, the noise level and the median eigenvalue of the covariance matrix are not necessarily linear, and consider the minimum eigenvalue instead. After all, the effectiveness of the PCA-PRI-PCAr algorithm is heavily based on accurately estimating the noise level. 
It should be noted that the noise map obtained by the NL-PCA filter requires Rician correction. \cite{manjon_mri_2015} used Monte Carlo simulation to obtain a mapping function for correcting the systematic noise underestimation. However, this method can be affected by the image itself, which means that the model is not universal. We can take a more general approach. The fixed point formula of SNR in \cite{koay_analytically_2006} establishes the analytical relationship between the magnitude signal-to-noise ratio and the correction factor. We can sample SNR at dense intervals to obtain the corresponding $\xi$ and $\gamma$ array, and then points of ($\gamma$, $\frac{1}{\sqrt{\xi}}$) are fitted to obtain a model closer to the analytical solution. The difference between the above two mapping functions is large at low SNR.
For the PRI-NLM filter, this paper suggests that modifying its weight to $w\left( i, j \right) =\exp \left[ -\frac{\left( g\left( i \right) -g\left( j \right) \right) ^2}{h_{i}^{2}} \right] $ can make it suitable for images with anisotropic voxels, where the best choice of parameter h is likely to change.

Good traditional algorithms are not overly dependent on the object being denoised. They should not be eliminated and can still be combined together or with deep learning methods in the future.  Additionally, we are interested in investigating the potential applications of extending our proposed PD tool and the PCA-PRI-PCAr algorithm to deal with image noise in spines, livers, or knees.

\section*{Acknowledgments}
\hspace{1em}

Data were provided in part by OASIS-1: Cross-Sectional: Principal Investigators: D. Marcus, R, Buckner, J, Csernansky J. Morris; P50 AG05681, P01 AG03991, P01 AG026276, R01 AG021910, P20 MH071616, U24 RR021382. Data were also provided by the Human Connectome Project, WU-Minn Consortium (Principal Investigators: David Van Essen and Kamil Ugurbil; 1U54MH091657) funded by the 16 NIH Institutes and Centers that support the NIH Blueprint for Neuroscience Research; and by the McDonnell Center for Systems Neuroscience at Washington University. This research did not receive any specific grant from funding agencies in the public, commercial, or
not-for-profit sectors.

\section*{Declaration of competing interests}
\hspace{1em}
None.

\bibliographystyle{apalike}

\bibliography{reference2}

\end{document}